\documentclass{sig-alternate-05-2015}

\usepackage{etoolbox}
\makeatletter
\patchcmd{\maketitle}{\@copyrightspace}{}{}{}
\makeatother

\usepackage{setspace}

\begin{document}

\CopyrightYear{2016} 
\setcopyright{rightsretained} 
\conferenceinfo{JCDL '16}{June 19-23, 2016, Newark, NJ, USA} 
\isbn{978-1-4503-4229-2/16/06}
\doi{http://dx.doi.org/10.1145/2910896.2925448}

%
% --- Author Metadata here ---
%\conferenceinfo{WOODSTOCK}{'97 El Paso, Texas USA}
%\CopyrightYear{2007} % Allows default copyright year (20XX) to be over-ridden - IF NEED BE.
%\crdata{0-12345-67-8/90/01}  % Allows default copyright data (0-89791-88-6/97/05) to be over-ridden - IF NEED BE.
% --- End of Author Metadata ---

\title{Semantometrics}
\subtitle{Towards Fulltext-based Research Evaluation}
%\titlenote{A full version of this paper is available as
%\textit{Author's Guide to Preparing ACM SIG Proceedings Using
%\LaTeX$2_\epsilon$\ and BibTeX} at

\numberofauthors{2} %  in this sample file, there are a *total*
\author{
% 1st. author
\alignauthor
Drahomira Herrmannova\\
%       \affaddr{KMi, The Open University}\\
       \affaddr{Milton Keynes, United Kingdom}\\
       \email{d.herrmannova@gmail.com}
% 2nd. author
\alignauthor
Petr Knoth\\
%       \affaddr{KMi, The Open University}\\
       \affaddr{Milton Keynes, United Kingdom}\\
       \email{petrknoth@gmail.com}
}
% There's nothing stopping you putting the seventh, eighth, etc.
% author on the opening page (as the 'third row') but we ask,
% for aesthetic reasons that you place these 'additional authors'
% in the \additional authors block, viz.
%\additionalauthors{Additional authors: John Smith (The Th{\o}rv{\"a}ld Group,
%email: {\texttt{jsmith@affiliation.org}}) and Julius P.~Kumquat
%(The Kumquat Consortium, email: {\texttt{jpkumquat@consortium.net}}).}
\date{\today}
% Just remember to make sure that the TOTAL number of authors
% is the number that will appear on the first page PLUS the
% number that will appear in the \additionalauthors section.

\maketitle

\begin{abstract}
Over the recent years, there has been a growing interest in developing new research evaluation methods that could go beyond the traditional citation-based metrics. This interest is motivated on one side by the wider availability or even emergence of new information evidencing research performance, such as article downloads, views and Twitter mentions, and on the other side by the continued frustrations and problems surrounding the application of purely citation-based metrics to evaluate research performance in practice. 

Semantometrics are a new class of research evaluation metrics which build on the premise that full-text is needed to assess the value of a publication. This paper reports on the analysis carried out with the aim to investigate the properties of the semantometric \textit{contribution} measure \cite{Knoth2014}, which uses semantic similarity of publications to estimate research contribution, and provides a comparative study of the contribution measure with traditional bibliometric measures based on citation counting.

%Our results suggest that while certain interesting relations can be found, the contribution measure might capture different aspects of performance than citation counts.
\end{abstract}




%
%  Use this command to print the description
%
%\printccsdesc

% We no longer use \terms command

%\keywords{ACM proceedings; \LaTeX; text tagging}
\vspace{-2mm}


\vspace{-1mm}
\section*{Acknowledgements}
This work was supported by Jisc under contract no. 3790.
\vspace{-2mm}

\section{Introduction}

% co jsou semantometrics a vysvetleni contribution metric

We have introduced the idea of Semantometrics in \cite{Knoth2014} as a new class of metrics for evaluating research. As opposed to existing Bibliometrics, Webometrics, Altmetrics, etc., Semantometrics are not based on measuring the number of interactions in the scholarly communication network, but build on the premise that full-text is needed to assess the value of a publication. 

In \cite{Knoth2014} we have attempted to create the first semantometric measure based on the idea of measuring the progress of scholarly discussion. Our hypothesis states that the added value of publication \textit{p} can be estimated based on the semantic distance from the publications cited by \textit{p} to the publications citing \textit{p}. This hypothesis is based on the process of how research builds on the existing knowledge in order to create new knowledge on which others can build. A publication, which in this way creates a ''bridge`` between what we already know and something new, which will people develop based on this knowledge, brings a contribution to science \cite{Knoth2014}.

Until recently, it was still technically challenging for us to obtain an evaluation dataset on which properties of the contribution metric could be analysed. In this respect, we are now able to report on the first large-scale analysis of this metric. The goal of our study was to understand the properties and behaviour of the semantometric contribution measure in comparison with established research evaluation metrics. We chose to use citation counts obtained from the Microsoft Academic Graph (MAG) \cite{Sinha2015}, as the representative of Bibliometrics, usage data (readership) obtained from Mendeley\footnote{\url{http://dev.mendeley.com}}, as the representative of Altmetrics and research articles aggregated by the Open Access Connecting Repositories\footnote{\url{http://core.ac.uk}} (CORE) system as a representative sample for studying the characteristics of the contribution measure. 

%The goal of this study is wasn't to advocate for the specific implementation of the contribution metric we originally suggested, but rather analyse it to see how it behaves. If improvements can be found, then we would like these to be implemented.

\vspace{-2mm}
\section{Dataset}

% jak jsme ziskali data, dataset statistics

\begin{figure*}[ht!]
    \centering
    \includegraphics[width=\textwidth]{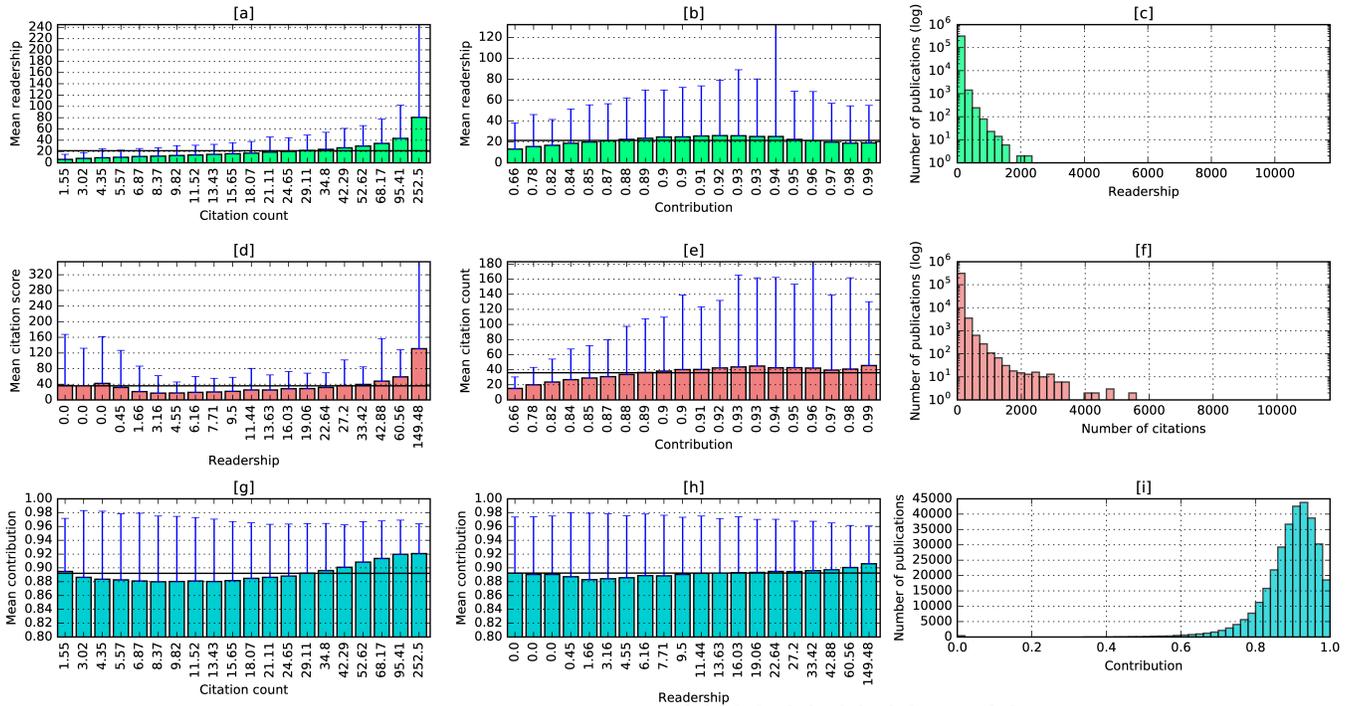}
    \vspace{-2.5em}
    \caption{Results of the study. To produce Figures [a], [b], [d], [e], [g] and [h], the data were split into 20 equally sized buckets by one of the studied metrics (x-axis). Mean and standard deviation of a second metric (y-axis) was then calculated for each of the buckets. The mean values are represented by the height of the bars, the vertical lines on top of the bars represent the standard deviations. The solid horizontal lines represents the mean value across all buckets.}
    % The mean values are represented by the height of the bars, the vertical lines on top of the bars represent the standard deviations.
    \label{fig:results}
    \vspace{-1.5em}
\end{figure*}

% For the purposes of our analysis we needed access to Open Access publications, their citation counts, reader counts and textual data of research papers citing and cited by the Open Access papers. To our knowledge such dataset did not exist, however we were able to create it by merging CORE, MAG and Mendeley datasets. 

Our experiments have been conducted on a dataset obtained by merging data from CORE, MAG and Mendeley. To assemble this dataset, we mapped DOIs of papers from CORE with MAG. Using MAG we then identified DOIs of papers citing and cited by the CORE papers. Finally, we used the DOIs to retrieve metadata, readership counts and primarily the titles and the abstracts using the Mendeley API. By merging these three datasets, we obtained a final dataset containing metadata, citation counts and reader counts of about 1.6 million Open Access papers. Additionally, we obtained metadata, including titles and abstracts of over 10 million papers which cite or are cited by the 1.6 million papers from CORE and are needed to calculate the contribution metric.

\vspace{-2mm}
\section{Results}

The main area of interest to us was the relation between the contribution measure and citation counts. The reason for this was the prevalence of use of citation counts in research evaluation. While using metrics based purely on citation counts has been subject to much criticism, these metrics still remain best known and most widely adopted. The aim was not to find a perfect correlation with citation counts, but rather demonstrate how does the contribution measure behave in relation to the well-known metric.

We have first investigated the distributions of the three metrics, these are shown in Figures \ref{fig:results} [c], \ref{fig:results} [f] and \ref{fig:results} [i]. As expected, the citation distribution (Figure \ref{fig:results} [f]) is a long tail (power law) distribution. This is consistent with existing studies \cite{Seglen1992}. The readership distribution (Figure \ref{fig:results} [c]) exhibits the same properties as the citation distribution. In contrast to the first two metrics, the contribution distribution (Figure \ref{fig:results} [i]) resembles a normal distribution.
%This has some implications, for example the skewness implies there will always be a large proportion of publications which are never cited, which makes it hard to evaluate and compare the impact of these publications among themselves. Furthermore, a power law distribution gives the impression that the majority of research outputs are of poor quality. 

%As opposed to the first two metrics,  A normal distribution is traditionally used to model the attainment of students, the performance of the workforce and also the ''quality`` of papers as measured in the peer-review system. As such it might also provide a better reflection of the distribution of research outputs' quality. On the other hand, some might argue that the normal distribution is not a true representation of the papers' (particularly economic or societal) impact.

To confirm our data are consistent with previous studies, we have investigated the relation between the citation and reader counts. We found that the two metrics are slightly correlated with Pearson $r = 0.3584$. A similarly strong correlation has been reported also by  \cite{schlogl2014downloads}. This correlation can also be seen when comparing the averaged values in Figures \ref{fig:results} [a] and \ref{fig:results} [d].

%To confirm our data are consistent with previous studies, we have investigated the relation between the citation and reader counts. We found that the two metrics are slightly correlated (Pearson $r=0.3584$). A similarly strong correlation was reported by \cite{schlogl2014downloads}. This correlation can also be seen when comparing the averaged values (Fig. \ref{fig:results} [a] and \ref{fig:results} [d]).

%In contrast to the previous case, we found no correlation between the contribution and citation counts (Pearson $r=0.0871$). However, we have made some interesting observations when comparing the averaged values (Figures \ref{fig:results} [e] and \ref{fig:results} [g]). Although the standard deviation shows it is not always the case, it seems publications with more than 25 citations are more likely to have higher contribution. However, once a paper receives around 90 citations, higher citation counts do not lead on average to a higher contribution. We think this is an interesting observation that is consistent with our perception of research quality. One possible and highly simplified explanation could be that receiving around 90 citations is typically an indication of quality work, especially when considering the size of certain research communities. Higher citation counts then typically reflect the size of the target audience community (impact) rather than the quality of the work.

In contrast to the reader counts, we found no correlation between the citation counts and contribution (Pearson $r=0.0871$). However, according to Figures \ref{fig:results} [g] and \ref{fig:results} [e] we can see that when comparing averaged values the behaviour of the contribution metric is not random, instead it is clearly correlated with citation counts. We can observe that publications with a citation score above a certain threshold achieve on average consistently higher contribution (Figure \ref{fig:results} [g]). Although the standard deviation shows it is not always the case, the results suggest that publications with more than 25 citations are more likely to have higher contribution. However, once a paper receives around 90 citations, higher citation counts do not lead on average to a higher contribution. We think this is an interesting observation that is consistent with our perception of research quality. One possible and highly simplified explanation could be that receiving around 90 citations is typically an indication of quality work. Higher citation counts then typically reflect the size of the target audience community (impact) rather than higher quality of the underlying research work. This leads us to the conclusion that the contribution metric seems to capture different aspects of research performance than citation counts.

%Furthermore, above a certain contribution value (~0.89) there are no differences in mean citation counts (Figure \ref{fig:results} [e]). It also seems that publications, whose contribution is below average, are less likely to be cited.

Similarly as in the previous case, there is no correlation between the contribution measure and reader counts, which is confirmed by Pearson $r=0.0444$. Interestingly, while we observed a correlation between the averaged contribution and citation counts, there seems to be no such relation between averaged contribution and reader counts (Figures \ref{fig:results} [b] and \ref{fig:results} [h]).

\vspace{-2mm}
\section{Conclusion}

%In this paper we presented a comparison of the semantometric contribution measure with citation counts and with Mendeley reader count. We observed some interesting properties of the contribution metric and relations between the contribution metric and citation counts. It seems that while the contribution metric seems to capture different aspects of research performance than citation counts, its behaviour is not random, but quite consistent and certain conclusions can be made based on the metric.

We have demonstrated that new measures for assessing publication impact, which take into account the manuscript of the publication, can be developed and presented a comparative study of the semantometric contribution measure with citation and reader counts. The results of our study suggest that the contribution metric captures different aspects of research performance than citation counts. More specifically, we believe that Semantometrics have the potential to capture research quality and contribution rather than research impact.

\vspace{-2mm}
\begin{spacing}{0.9}
\bibliographystyle{abbrv}
{\bibliography{bibliography}}

\begin{thebibliography}{1}

\bibitem{Knoth2014}
P.~Knoth and D.~Herrmannova.
\newblock {Towards Semantometrics: A New Semantic Similarity Based Measure for
  Assessing a Research Publication's Contribution}.
\newblock {\em D-Lib Magazine}, 20(11/12), 2014.

\bibitem{schlogl2014downloads}
C.~Schl{\"o}gl, J.~Gorraiz, C.~Gumpenberger, K.~Jack, and P.~Kraker.
\newblock Are downloads and readership data a substitute for citations? the
  case of a scholarly journal.
\newblock {\em LIDA Proceedings}, 13, 2014.

\bibitem{Seglen1992}
P.~O. Seglen.
\newblock {The Skewness of Science}.
\newblock {\em JASIS}, 43(9):628--638, oct 1992.

\bibitem{Sinha2015}
A.~Sinha, Z.~Shen, Y.~Song, H.~Ma, D.~Eide, B.-j.~P. Hsu, and K.~Wang.
\newblock {An Overview of Microsoft Academic Service (MAS) and Applications}.
\newblock In {\em Proceedings of WWW 2015}, pages 243--246, Florence, Italy,
  2015. ACM Press.

\end{thebibliography}
\end{spacing}
%\bibliographystyle{abbrv}
%\bibliography{bibliography}  % sigproc.bib is the name of the Bibliography in this case
% ACM needs 'a single self-contained file'!
%\balancecolumns
%\appendix
%Appendix A
%\balancecolumns % GM June 2007
% That's all folks!
\end{document}